\documentclass[twocolumn,aps,tightenlines,floatfix,showpacs]{revtex4}
\usepackage{graphicx}
\begin{document}

\title{Constraints on the Skyrme Equations of State from Properties
of Doubly Magic Nuclei}

\author{B. Alex Brown}

\affiliation{National Superconducting Cyclotron Laboratory and Department
of Physics and Astronomy, Michigan state University, East Lansing,
Michigan 48824-1321, USA}

\begin{abstract}
I use properties of doubly-magic nuclei to
constrain nuclear matter and neutron matter equations of state.
I conclude that the data determined the
value of the neutron equation of state and the symmetry energy
near a density of $\rho_{on}$ = 0.10 nucleons/fm$^{3}$.
The slope at that point
is constrained by the value of the neutron skin.
\end{abstract}


\maketitle

The neutron equation of state (EOS)
is important for understanding
properties of neutron stars \cite{steiner13}, \cite{steiner05}.
Recently an extensive study was made of the constraints
on the Skyrme energy-density functionals (EDFs) provided by the properties
of nuclear matter \cite{dutra}. Out of several hundred
Skyrme EDFs, the 16 given in Table VI of \cite{dutra} called
the CSkP set best reproduced
a selected set of experimental nuclear matter properties.
Five of these were eliminated \cite{dutra} since
they gave transitions to spin-ordered matter
around $\rho$ = 0.25 nucleons/fm$^{3}$ (in this
paper the densities are all in units of nucleons/fm$^{3}$).
The remaining 11 are those given in Table I
and labeled with their name and order in Table VI of \cite{dutra}.
I start with these 11 interactions and use properties
of doubly-magic nuclei to provide further constraints on the deduced EOS.
I also consider the SLy4 \cite{sly4} and SkM* \cite{skms} EDFs since they
are widely used in the literature.
\begin{figure}
\includegraphics[scale=0.4]{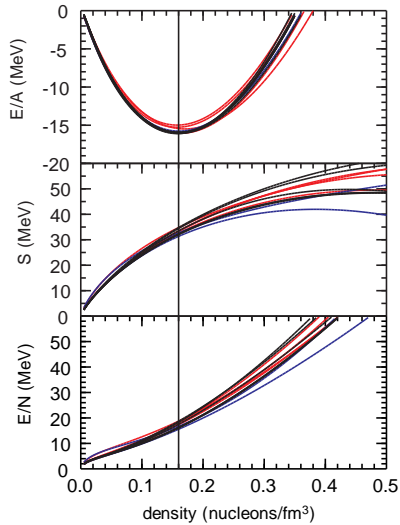}
\caption{EOS obtained from the CSkP set of Skyrme interactions plus
SLy4 and SkM*. The black lines
are those with m$^{*}$/m $  \approx  $ 1.0 and the red lines are those
with m$^{*}$/m = 0.70-0.85. The blue lines are those for SLy4 and SkM*.}
\end{figure}
\begin{figure}
\includegraphics[scale=0.4]{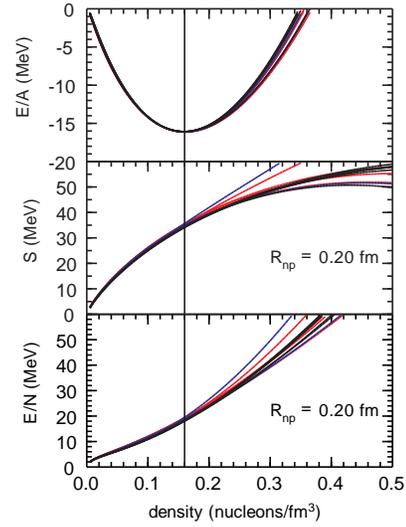}
\caption{EOS obtained from the CSkP set of Skyrme interactions
fitted to properties of doubly-magic nuclei and with a constraint
of $  R_{np} = 0.20  $ fm for the neutron skin of $^{208}$Pb.
See caption to Fig. 1.}
\end{figure}
\begin{figure}
\includegraphics[scale=0.4]{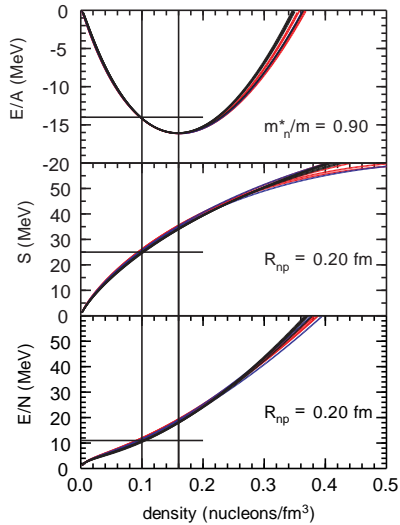}
\caption{EOS obtained from the CSkP set of Skyrme interactions
fitted to properties of doubly-magic nuclei and with a constraint
of 0.20 fm for the neutron skin of $^{208}$Pb and m$^{*}_{n}$/m=0.90 at
$\rho_{on}$ = 0.10.  See caption to Fig. 1.}
\end{figure}
\begin{figure}
\includegraphics[scale=0.4]{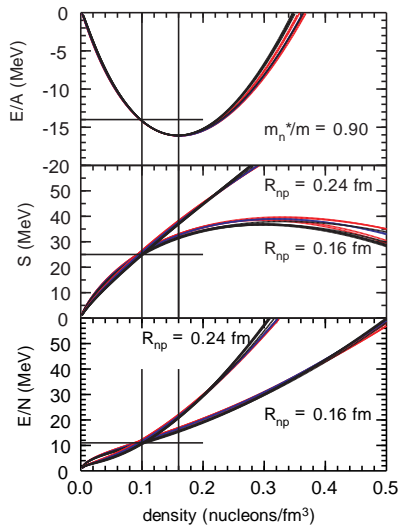}
\caption{EOS obtained from the interactions fitted
to properties of doubly-magic nuclei and with values of
0.16 and 0.24 fm for the neutron
skin of $^{208}$Pb together with m$^{*}_{n}$/m=0.90. See caption to Fig. 1.}
\end{figure}

The EOS
for nuclear matter, neutron matter and the symmetry energy $  S  $
obtained with the CSkP, SLy4 and SkM* parameters
given Table VI of \cite{dutra} are shown in Fig. 1. The symmetry energy EOS is
the difference between the neutron EOS and matter EOS. The spread is
relatively small. But I will demonstrate that the spread can be reduced if the
Skyrme parameters are constrained by the same set of nuclear data.

The data set consists of the ground-state properties of the doubly-magic nuclei
used for
the Skx family of Skyrme interactions \cite{skx}, \cite{skxb}, \cite{skxc},
\cite{skxd}, \cite{skxe}: $^{16}$O, $^{24}$O, $^{34}$Si
$^{40}$Ca, $^{48}$Ca, $^{48}$Ni, $^{68}$Ni, $^{88}$Sr, $^{100}$Sn, $^{132}$Sn and $^{208}$Pb.
The properties are binding energies, rms charge radii and single-particle
energies. These data are given in \cite{skx}.
All of the CSkP functionals were constrained to have a range of
 $  L  $ values (defined below) that correspond to a neutron
skin thickness (the differences between the neutron and proton rms radii)
for $^{208}$Pb to be near $  R_{np} = 0.20  $ fm. The
The Ska25s20 and Ska35s20 are unpublished functionals in the Skx family
that were constrained to have $  R_{np} = 0.20  $ fm. I start
by constraining the skin thickness to be exactly
$  R_{np} = 0.20  $ fm. This constraint will later be relaxed.
The Skx family drops the Coulomb exchange term in order to
reproduce the binding energy differences between $^{48}$Ni and $^{48}$Ca \cite{skxb}.

Starting with the parameters  given in Table VI of \cite{dutra}
I refit $  t_{0}, t_{1}, t_{2}, t_{2}, x_{0}  $ and $  x_{3}  $ to the doubly-magic data for each 
case.
The functional forms for the Skyrme EDF and its nuclear
matter properties are given in \cite{dutra}.
These parameters are all well determined by the data set.
The values for $\sigma$, $  x_{1}  $ and $  x_{2}  $ were fixed at their original
values. A reasonable range of $\sigma$ gives relatively good fits to the data.
The $  x_{1}  $ and $  x_{2}  $ are not determined from the data.
The parameters that are fixed are given in Table I together with the
value of the $\chi^{2}$ of the fit. The 4th one in Table VI of \cite{dutra}
(LNS) was not used since it did not give converged results for nuclei.
The new results for the EOS are shown in Fig. 2. The EOS are now more
tightly constrained.
The single-particle energies included in the fit
are best reproduced with a nuclear-matter effective mass of near unity.
The fit quickly converges for the values of the binding energies
and radii. The convergence for single-particle energies is an order
of magnitude slower. If all of the fits were fully converged, they
would end up with the same low $\chi^{2}$ values and with
nuclear matter effective masses near unity. I stop with the
quick convergence part that leaves some range of values for
the nuclear matter effective mass.
\begin{figure}
\includegraphics[scale=0.4]{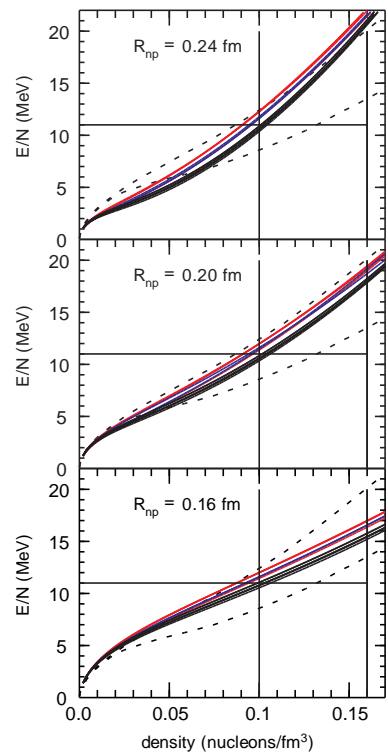}
\caption{The neutron EOS at low density for $  R_{np}  $ = 0.16, 0.20 and 0.24 fm.
They are compared to the theoretical results from \cite{heb} (dashed lines).
See caption to Fig. 1.}
\end{figure}

Some of the CSkP set were eliminated in \cite{dutra} because the neutron matter
effective mass was not less than unity. However, I find that
by repeating the fit and allowed the $  x_{1}  $ parameter to vary
and keeping $  x_{2}  $ fixed,
the neutron effective mass at $\rho_{on}$=0.10 can be fixed at a value of 0.90
without any significant change to the $\chi^{2}$ for the fit to nuclear
ground-state data. (The neutron effective mass depends upon both
$  x_{1}  $ and $  x_{2}  $. In some cases $  x_{2}  $ and be varied and
$  x_{1}  $ can be fixed. In the case of Sly4 it was only
possible to do the latter.)
The EOS are shown in Fig. 3. They are similar to those
in Fig. 2, but are closer together at large density.
The resulting $  x_{1}  $ values are given on
the right hand side of Table I. A good fit would be obtained
for a reasonably wide range of values for the neutron
effective mass at $\rho_{on}$=0.10 (0.8 to 1.0).

The CSkP sets provide a reasonable range of values for nuclear matter
incompressibility $  K_{m}  $ from 219 to 240 MeV as given in Table I.
The fit to nuclear ground-state data is good for all of this range.
The range of values obtained from a QRPA analysis of the
giant monopole resonance is 217-230 MeV \cite{sag}.

The fits are then performed with constraints of $  R_{np}=0.16  $
and $  R_{np}=0.24  $ fm. The results are shown in Fig. 4.
The $\chi^{2}$ values are nearly the same for this range of $  R_{np}  $.
The neutron EOS for these two fits cross at values
from $\rho$=0.096 to $\rho$=0.104 indicating that the value of the neutron
EOS is most well determined in this range. I take
the average value of $\rho_{on}$=0.10 to define the point at which
the value of the neutron EOS is most well determined.
This crossing point is the same as that found in \cite{skin}.
The crossing point depends a little on the neutron effective mass;
the average would be closer to $\rho$=0.11 if the neutron effective mass
was unity. The average neutron density for
the interior of $^{208}$Pb is about $\rho_{n}$ = (126/208)($\rho_{om}$) = 0.097
with $\rho_{om}$ = 0.16.
The values of the neutron EOS at $\rho_{on}$=0.10 range from 10.5 to 12.2 MeV
with a mean of 11.3(8) MeV. If we take only the best fit results with
a nuclear matter effective mass near unity them the mean is 10.8(3) MeV.
\begin{table*}
\begin{center}
\caption{Properties of the fitted Skyrme interactions with $  R_{np} = 0.20  $ fm. The results for
$  x_{1}  $ in the last column were obtained with the additional
constraint that $  m^{*}_{n}(\rho _{on})/m \approx 0.90  $.}
\begin{tabular}{|c|c||c|c|c|c|c|c|c||c|}
\hline
   name &      & $\sigma$    & $  x_{1}  $ & $  x_{2}  $&  $\chi^{2}$ & $  K_{m}  $ & m$^{*}_{n}$/m 
 & m$^{*}$/m   & $  x_{1}  $  \\
        &      &       &         &        &       &  (MeV) &  at $\rho_{on}$=0.10
        & at $\rho_{om}$=0.16 & \\
\hline
KDE0v1   & s3  &  1/6  & -0.35  &  -0.93 & 1.88  & 219 & 0.79 & 0.79 & 0.18  \\
LNS      & s4  &  1/6  &  0.06  &  0.66  &       &     &      &      &        \\
NRAPR    & s6  & 0.14  & -0.05  &  0.03  & 2.77  & 227 & 1.00 & 0.85 & -0.07  \\
Ska25s20 & s7  & 0.25  & -0.80  &  0.00  & 0.88  & 219 & 0.99 & 0.99 & -1.35  \\
Ska35s20 & s8  & 0.35  & -0.80  &  0.00  & 0.74  & 240 & 1.00 & 1.00 & -1.46  \\
SKRA     & s9  & 0.14  &  0.00  &  0.20  & 1.70  & 215 & 0.99 & 0.79 & -0.45  \\
SkT1     & s10 & 1/3   & -0.50  & -0.50  & 0.79  & 236 & 0.98 & 0.97 & -1.02  \\
SkT2     & s11 & 1/3   & -0.50  & -0.50  & 0.82  & 237 & 0.95 & 0.97 & -0.99  \\
SkT3     & s12 & 1/3   & -1.00  &  1.00  & 0.76  & 236 & 0.98 & 0.98 & -1.52  \\
SQMC750  & s15 & 1/6   &  0.00  &  0.00  & 2.50  & 227 & 0.92 & 0.71 & -0.08  \\
SV-sym32 & s16 & 0.30  & -0.59  & -2.17  & 0.83  & 233 & 1.12 & 0.91 & -1.77  \\
\hline
SLy4     & s17 & 1/6   & -0.34  & -1.00  & 3.42  & 230 & 0.73 & 0.70 &        \\
SkM*     & s18 & 1/6   &  0.00  &  0.00  & 1.71  & 221 & 0.98 & 0.78 & -0.36  \\
\hline
\end{tabular}
\end{center}
\end{table*}
\begin{table*}
\begin{center}
\caption{Nuclear matter properties of the Ska25 fits. $  J  $, $  L  $ and $  K  $ are
defined at $\rho_{om}$=0.16. $  F(\rho _{on})  $ and $  F'(\rho _{on})  $ are defined at 
$\rho_{on}$=0.10. }
\begin{tabular}{|c|c||c|c|c|c||c|c|c||c|c|}
\hline
  $  R_{np}  $ & & $  a  $ & $  b  $ & $  c  $ & $  d  $ & $  J  $ & $  L  $ & $  K  $ & $  F(\rho 
_{on})  $ & $  F'(\rho _{on})  $ \\
\hline
     & $  F_{m} = E/A  $   & -818 &  913  &  75 &   8 & -16.0 & -0.1   &  219 &  -14.1 & -66 \\
\hline
0.16 & $  F_{n} = E/N  $   & -307 &  238  & 119 & 117 &  15.6 & 40.7 &   53 &   10.8   &   73 \\
     & $  F_{sym} = S  $ &  511 & -675  &  44 & 116 &  31.9 & 40.8 & -166 &   25.0   &  139 \\
\hline
0.20 & $  F_{n} = E/N  $   & -444 &  475  & 119 & 124 &  17.9 & 66.5 &  123 &   10.6   &  105 \\
     & $  F_{sym} = S  $ &  374 & -438  &  44 & 116 &  34.0 & 66.6 &  -96 &   24.7   &  171 \\

\hline
0.24 & $  F_{n} = E/N  $   & -568 & 695  & 119 & 133 &  20.8 &  92.6 &  190 &   10.8   &  139 \\
     & $  F_{sym} = S  $ &  250 & -218 &  44 & 116 &  36.4 &  92.7 &  -29 &   24.9   &  205 \\
\hline
\end{tabular}
\end{center}
\end{table*}

The low-density results for the neutron EOS are shown in Fig. 5
for the three values of $  R_{np}  $.
They are compared with the upper and lower values of
the error band of calculations based on the Entem and Machleidt
N$^{3}$LO NN potential with a cutoff at 500 MeV that
include N$^{2}$LO NNN forces from \cite{heb} (the dashed lines).
Other predictions are
shown in Fig. 7 of \cite{kru}.
The present results for $  R_{np} = 0.20  $ fm
are all within the theoretical error band and have the
same $\rho$ dependence as the calculations.
The present results will
be important for comparing to future calculations with smaller
error bands.

All of the curves obtained by Skyrme EDF in these figures are given by
the analytical expression
$$
F(\rho ) =a \rho  + b \rho ^{\gamma } + c \rho ^{2/3} + d \rho ^{5/3},       \eqno({1})
$$
where $  \gamma  = 1 + \sigma   $, and $  a, b, c  $ and $  d  $ are constants that depend
on the Skyrme parameters. The first term is from the delta-function part
that depends on $  t_{0}  $ and $  x_{0}  $,
the second term is from the density dependent part that depends on
$  t_{3}  $ and $  x_{3}  $,
the third term is the Fermi-gas kinetic energy, and the fourth term
depends on $  t_{1}, t_{2}, x_{1}  $ and $  x_{2}  $. The effective mass is given by
$$
\frac{m^{*}(\rho )}{m} = \frac{c}{c+d\rho }.       \eqno({2})
$$
The first and second derivatives are given by
$$
F'(\rho ) =a + b\gamma  \rho ^{\gamma -1} + (2/3) c \rho ^{-1/3} + (5/3) d \rho ^{2/3},       
\eqno({3})
$$
and
$$
F''(\rho ) = \gamma (\gamma -1)b \rho ^{\gamma -2} - (2/9) c \rho ^{-4/3} + (10/9) d  \rho ^{-1/3}. 
      \eqno({4})
$$
Given a fixed $  c  $ and $  d  $ one can write
$$
[F(\rho )/\rho ] = a + b \rho ^{\gamma -1} + A(\rho )       \eqno({5})
$$
and
$$
F'(\rho ) = a + b \gamma  \rho ^{\gamma -1} + B(\rho ),       \eqno({6})
$$
where
$$
A(\rho ) = c \rho ^{-1/3} + d \rho ^{2/3},       \eqno({7})
$$
and
$$
B(\rho ) = (2/3) c \rho ^{-1/3} + (5/3) d \rho ^{2/3}.       \eqno({8})
$$

These equations provide the results needed
to obtain $  a  $ and $  b  $ in terms of $  F(\rho _{o})  $
and $  F'(\rho _{o})  $ at a fixed value $\rho$=$\rho_{o}  ,  $
$$
b = \frac{F'(\rho _{o}) - [F(\rho _{o})/\rho _{o}] - A(\rho _{o}) + B(\rho _{o})}{(\gamma -1)\rho 
_{o}^{\gamma -1}},       \eqno({9})
$$
and
$$
a = [F(\rho _{o})/\rho _{o}] - b \rho _{o}^{\gamma -1} - A(\rho _{o}).       \eqno({10})
$$

It is conventional to define the value and derivatives at $\rho_{om}$ = 0.16
with $  J = F(\rho _{om})  $, $  L = 3 \rho _{om} F'(\rho _{om})  $ and
$  K = 9 \rho _{om}^{2} F''(\rho _{om})  $.
There are three quantities, the nuclear-matter EOS, $  F_{m} = (E/A)  $,
the neutron EOS, $  F_{n} = (E/N)  $, and the symmetry energy
$  S = F_{sym} = F_{n} - F_{m}  $.
The results corresponding to one of the best fits to nuclear ground-state
data, the Ska25 with $\sigma$=0.25, are given Table II.
For a given $\gamma$ and effective mass ($  d  $),
the values of $  F(\rho _{o})  $ and $  F'(\rho _{o})  $ determine the entire
EOS. For example, for
the nuclear matter EOS the values $  F(\rho _{om})=-16.0  $, $  F'(\rho _{om})=0.0  $,
$  c=75  $ and $  d=0  $
determine $  a  $ and $  b  $ and the entire nuclear matter EOS.

Considering the entire CSkP set,
the neutron EOS is best determined at $\rho_{on}$ = 0.10
with a value of $  F_{n}(\rho _{on})  = (E/N)(\rho _{on}) =  11.3(8)  $
MeV. The symmetry energy at this point
is $  S(\rho _{on}) =  F_{n}(\rho _{on}) - F_{m}(\rho _{on})  $
= 11.3(8) $+$ 14.1(1) = 25.4(8) MeV.
The derivatives at $\rho_{on}$=0.10 are approximately
linear with the value of $  R_{np}  $ with
$$
S'(\rho _{on}) = p_{s} R_{np}.       \eqno({11})
$$
and
$$
F'_{n}(\rho _{on}) = p_{n} R_{np}.       \eqno({12})
$$
with $  p_{s}=850  $ and $  p_{n}=525  $.
For a given value of $  R_{np}  $ we have $  S'  $ and $  F'_{n}  $
that can be used together with the above values at $\rho_{on}$=0.10
in Eqs. (9) and (10) to obtain
$  a  $ and $  b  $ and the entire neutron EOS.
The equations above provide analytical forms for the
correlations between $  a  $, $  b  $, $  d  $, $\gamma$, $  J  $, $  L  $,
$  K  $, $  F_{n}(\rho _{on})  $,  $  F'_{n}(\rho _{on})  $ and $  R_{np}  $.
Although the $\gamma$=1$+\sigma$ value
and the $  d  $ term are not the
same for all of the Skyrme EDFs considered, all of the lines in the Fig. 3, 4 and 5
are within a narrow band that includes Ska25 lines. Thus the parameters given
in Table II are close to a universal parametrization
of the Skyrme EOS that is constrained by nuclear data and its dependence on the neutron skin.
An assumption for all of the Skyrme EDFs considered is that
$  \gamma =1+\sigma   $ is the same for both the matter and neutron EOS.
The $\gamma$ for nuclear matter is constrained by the $  K_{m}  $ value
obtained from the energy of the giant monopole resonance.
The value of $\gamma$ may be different for neutron matter and this would effect
the $  K_{sym}  $ value. The $^{208}$Pb neutron skin thickness
obtained from the PREX parity-violating
electron scattering experiment
is $  R_{np}=0.302\pm(0.175)_{{\rm exp}}\pm(0.026)_{{\rm model}}\pm(0.005)_{{\rm strange}}  $
fm \cite{prex}, \cite{hor}.
A PREX-II experiment has been approved that will
reduce the error bar to about 0.06 fm. This will put an important constraint
on the neutron EOS. Other data that will constrain the neutron EOS come
from properties of the dipole resonance \cite{dipole1}, \cite{dipole2} and from nuclear
reactions \cite{reaction}.

{\bf Acknowledgements:}
I acknowledge support from NSF grant PHY-1068217.

\end{document}